\documentclass[
]{ceurart}

\usepackage[frozencache]{minted}

\makeatletter
\AtBeginEnvironment{minted}{\dontdofcolorbox}
\def\dontdofcolorbox{\renewcommand\fcolorbox[4][]{##4}}
\makeatother

\setminted{breaklines=true}

\begin{document}

\copyrightyear{2025}
\copyrightclause{Copyright for this paper by its authors. Use permitted under Creative Commons License Attribution 4.0 International (CC BY 4.0).}
\conference{ISWC 2025 Companion Volume, November 2--6, 2025, Nara, Japan}

\title{Jelly-Patch: a Fast Format for Recording Changes in RDF Datasets}

\author[1,2]{Piotr Sowiński}[%
orcid=0000-0002-2543-9461,
email=piotr@neverblink.eu,
url=https://ostrzyciel.eu/,
]
\cormark[1]
\address[1]{NeverBlink, ul. Wspólna 56, 00-684 Warsaw, Poland}
\address[2]{Warsaw University of Technology, Pl. Politechniki 1, 00-661 Warsaw, Poland}

\author[1]{Kacper Grzymkowski}[%
orcid=0009-0008-9227-8240,
email=kacper.grzymkowski@neverblink.eu,
]

\author[1,2]{Anastasiya Danilenka}[%
orcid=0000-0002-3080-0303,
email=anastasiya@neverblink.eu,
]

\cortext[1]{Corresponding author.}

\begin{abstract}
    Recording data changes in RDF systems is a crucial capability, needed to support auditing, incremental backups, database replication, and event-driven workflows. In large-scale and low-latency RDF applications, the high volume and frequency of updates can cause performance bottlenecks in the serialization and transmission of changes. To alleviate this, we propose Jelly-Patch -- a high-performance, compressed binary serialization format for changes in RDF datasets. To evaluate its performance, we benchmark Jelly-Patch against existing RDF Patch formats, using two datasets representing different use cases (change data capture and IoT streams). Jelly-Patch is shown to achieve 3.5--8.9x better compression, and up to 2.5x and 4.6x higher throughput in serialization and parsing, respectively. These significant advancements in throughput and compression are expected to improve the performance of large-scale and low-latency RDF systems.
\end{abstract}

\begin{keywords}
  RDF \sep
  Change data capture \sep
  Diffs \sep
  RDF Patch \sep
  Databases \sep
  Serialization format
\end{keywords}

\maketitle

\section{Introduction}

Recording changes in a dataset or tracking ``diffs'' (deltas) between datasets is a common problem in modern data-intensive systems~\cite{kleppmann2017designing}, including RDF databases. For instance, change data capture (CDC) is a popular pattern where all changes made to a database are written to a log or a stream. This log can be used for auditing, incremental backups, replication in high-availability setups, or implementing downstream event-driven systems. For RDF databases, the problem boils down to tracking statements that were added and removed in each transaction~\cite{mynarz2019change}.

Several solutions for RDF change tracking were proposed~\cite{berners2004delta}, including LD Patch~\cite{Sambra:15:LDP}, SparqlPatch~\cite{web:sparqlpatch}, and RDF Patch~\cite{web:rdfdelta} (implemented in Apache Jena). One of the key issues for these formats was updating triples that contain blank nodes, with the first two adopting elaborate update operators that preserve the anonymity of blank nodes. RDF Patch takes a simpler approach, where blank node identifiers are treated as global and unique. In other words, a single blank node identifier in two different RDF documents identifies the same blank node. While this deviates from the RDF specification, it is a practical choice for database use cases and CDC patterns that are common in the industry, greatly simplifying the implementation. RDF Patch was adopted as the basis for the RDF Delta replication system, used in High Availability Apache Jena Fuseki~\cite{web:rdfdelta}. It uses a text-based serialization format based on N-Quads, with additional markers indicating add/delete operations and transactions:

\begin{minted}{turtle}
TX .  # Transaction begin
A _:sensor001 <http://example.org/hasTemperature> "23" .  # Add triple
D _:sensor001 <http://example.org/hasTemperature> "22" .  # Delete triple
TC .  # Transaction commit
\end{minted}

Large-scale RDF databases and low-latency streaming applications require the highest levels of performance from serialization formats, so as not to bottleneck the system with slow serialization. This includes OLTP databases working with very frequent or large transactions, which must record the changes efficiently to quickly close each transaction. In streaming applications, small incremental changes to RDF datasets are common, such as the sensor value update in the example above. In this case, reducing the serialization/deserialization time and the size of the representation will result in lower latency and decreased usage of computing resources.

To answer these needs, in this work we introduce \textbf{Jelly-Patch}, an efficient binary serialization format for RDF Patch. It is based on Jelly-RDF, a high-performance RDF serialization format~\cite{sowinski2025jelly}. Jelly-Patch was designed as a faster and more compressed alternative to the RDF Patch text format, aiming to improve the scalability and responsiveness of RDF systems. Our contributions include: (1) an open specification of the Jelly-Patch format; (2) open-source implementation in Java, integrated with Apache Jena and RDF4J; and (3) performance evaluation comparing Jelly-Patch to alternative formats.

\section{Jelly-Patch} \label{sec:format}

Jelly-Patch uses Protocol Buffers, a fast and widely used binary serialization framework~\cite{web:protobuf}. Jelly-Patch reuses the base message types for triples, quads, IRIs, literals, etc. from Jelly-RDF. On top of these, RDF Patch-specific messages were added for transactions, patch headers, and support for adding/deleting prefixes. Jelly-Patch is defined in an open specification,\footnote{\url{https://w3id.org/jelly/dev/specification/patch}} accompanied by an interface definition file that can be used to generate serialization/deserialization code in any popular programming language. The format covers the entirety of RDF 1.1, RDF-star, and generalized RDF.

The base compression scheme in Jelly-Patch is the same as in Jelly-RDF~\cite{sowinski2025jelly,sowinski2022efficient}, making it possible to reuse much of the same code between the two formats. The key difference between the two formats is that while Jelly-RDF expresses a stream of statements, Jelly-Patch describes a stream of changes, which can be thought of as a derivative of a stream. A single Jelly-Patch stream can contain many patches. Compression works in a fully streaming manner over the entire stream, which is especially advantageous for CDC workloads. For example, if a given IRI is present in a patch earlier in the stream, later patches can refer to the same IRI through a streaming lookup table, reducing file size.

We implemented Jelly-Patch in Java, as part of the Jelly-JVM library,\footnote{\url{https://w3id.org/jelly/jelly-jvm/}} licensed under Apache 2.0. The core serialization code is generic and can be integrated with any RDF library for the Java Virtual Machine. We integrated it fully with Apache Jena, including high-level APIs on par with Jena's internal serialization formats. We also implemented a low-level integration with RDF4J, which is limited due to RDF4J not supporting RDF Patch natively.

\section{Evaluation}

We evaluated Jelly-Patch in terms of its serialized representation size, serialization throughput, and deserialization throughput. The benchmarks were performed with two datasets, representing very different use cases: change data capture of an RDF database, and streaming IoT sensor data.

\subsection{Datasets}

The change data capture dataset (\texttt{bsbm-cdc}) was created using the Berlin SPARQL Benchmark (BSBM)~\cite{Bizer2009bsbm} test driver, running the \texttt{update} workload against the RDF Delta server 1.1.2~\cite{web:rdfdelta}. The BSBM data generator was configured with a scale factor of 100,000, and a transaction count of 100,000. The BSBM test driver executed 90,000 query mixes, each consisting of 2 insert queries and 3 delete queries. A single-node RDF Delta setup was used to capture the changes to RDF Patch files. The resulting patches were then combined into a single dataset.

The streaming IoT sensor dataset (\texttt{assist-iot-weather}) is based on 
a RiverBench~\cite{sowinski2024realizing} dataset,\footnote{\url{https://w3id.org/riverbench/datasets/assist-iot-weather-graphs/1.0.3}} which consists of sensor readings from an IoT weather station. It was processed by calculating the rolling difference between consecutive timestamped graphs and writing the differences as patches.

Table~\ref{tab:datasets} presents key statistics about both datasets. In this case, we consider a patch to correspond to exactly one transaction. An operation is a single row in an RDF Patch file (e.g., transaction begin, triple add/delete, header). The datasets are publicly available on Zenodo under the CC BY 4.0 license~\cite{zenodo}.

\begin{table}[htb]
    \caption{Summary of the datasets used in benchmarks.}
    \label{tab:datasets}
    \centering
    \begin{tabular}{lrrrrr}
        \toprule
        \textbf{Dataset} & \textbf{Patches} & \textbf{Operations} & \textbf{Triple adds} & \textbf{Triple deletes} & \textbf{Size in RDF Patch} \\
        \midrule
        \texttt{bsbm-cdc} & 450,448 & 38,368,364 & 34,712,608 & 1,853,960 & 8914\,MB \\
        \texttt{assist-iot-weather} & 701,277 & 17,032,881 & 7,815,221 & 7,815,106 & 2499\,MB \\
        \bottomrule
    \end{tabular}
\end{table}

\subsection{Benchmark Setup}

Benchmarks were implemented on top of Apache Jena 5.3.0, using Jelly-JVM 3.4.0. We used the following Jelly-Patch settings: name table size 4000, prefix table size 1024, frame size 512. We compared it against formats built into Jena: RDF Patch text (same as described in the RDF Patch specification), RDF Patch binary (based on Jena's binary Thrift format~\cite{web:jena-binary}), and SPARQL Update (INSERT/DELETE DATA). All tested methods are based on the same underlying Jena APIs, programming language, and execution environment, to make the comparison as fair as possible. Similarly to Jelly-Patch, the RDF Patch formats built into Jena are based on the already heavily optimized implementations of N-Quads and Jena Binary formats that are used in production environments.

For serialization, an in-memory list of changes was sent to the serializer writing to a blackhole output stream, discarding the written bytes. For deserialization, the parser was reading from an in-memory byte array, containing pre-serialized data. The results were sent to a blackhole change handler.

The benchmark was implemented using the Java Microbenchmark Harness (JMH) 1.37, an industry-standard benchmarking tool that accounts for JVM warmup and dead-code elimination using JVM blackholes, ensuring fair results~\cite{web:jmh}. JMH was run with parameters: \texttt{-f7 -wi 10 -i 20 -gc true} in single-shot mode. Test bench used: AMD Threadripper 7960X (24-core, 4.2 GHz); 256 GB RAM (DDR5 4800 MHz); Linux 6.8.0; Oracle GraalVM 24.0.2+11.1. The disk was not used during the benchmarks. Benchmarks were single-threaded, but the JVM was allowed to use the remaining threads for garbage collection and JIT compilation. Full benchmark code and scripts can be found on GitHub\footnote{\url{https://github.com/Jelly-RDF/jvm-benchmarks}}.

\subsection{Results}

\begin{figure}[htbp]
    \centerline{\includegraphics[width=\textwidth]{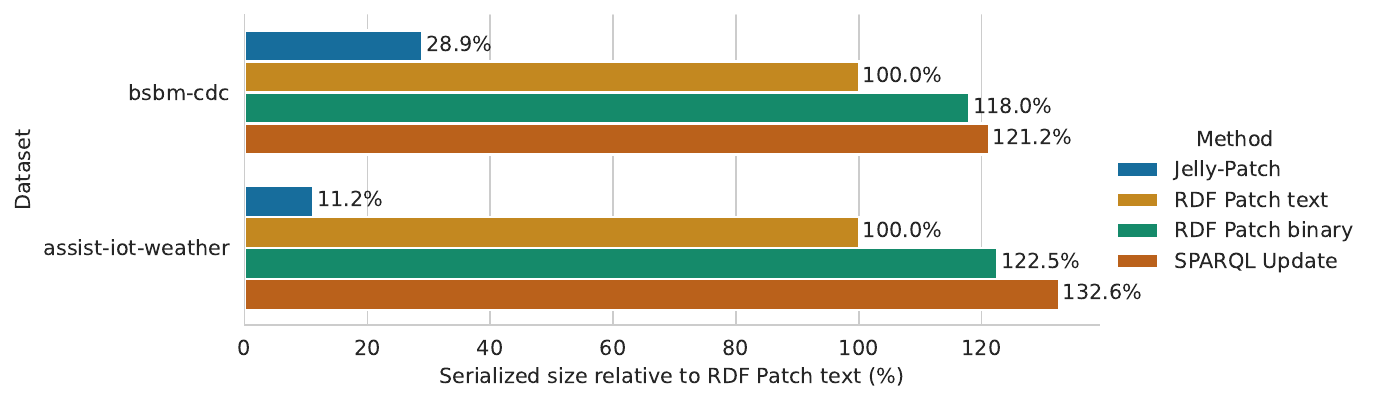}}
    \caption{Serialized dataset size, expressed as percentage of baseline (dataset's size in RDF Patch text).}
    \label{fig:size}
\end{figure}

Figure~\ref{fig:size} compares the serialized representation size of the four formats, with RDF Patch text being used as the baseline. RDF Patch binary does not employ any compression, and results in files $\sim$20\% larger than with the text format. Jena's SPARQL Update implementation also does not use any compression, with files 21--33\% larger than the text format. Jelly-Patch is the only format employing any compression, reducing the file size by 3.5x for \texttt{bsbm-cdc}, and by \textbf{8.9x} for \texttt{assist-iot-weather}.

The observed difference in compression ratios is due to \texttt{bsbm-cdc} using long literals and a very large pool of quickly changing IRIs (e.g., 100,000 products). Jelly-Patch does not apply binary compression to literals -- to alleviate this, the file would have to be additionally compressed with, for example, gzip. Conversely, in \texttt{assist-iot-weather} the literals are small and IRIs repeat often, making them prone to Jelly's compression mechanisms.

The compressed weather dataset when saved as a change stream in Jelly-Patch takes up only 279.6 MB, as compared to 1513.8 MB of the original RiverBench dataset saved in Jelly-RDF, while not losing any information. This is a size reduction of \textbf{5.4x}, made possible by applying differential compression (stream derivative) on top of an already well-compressed file. This highlights the great potential of diff-based formats in streaming use cases, which we will investigate further in future research.

\begin{figure}[htbp]
    \centerline{\includegraphics[width=\textwidth]{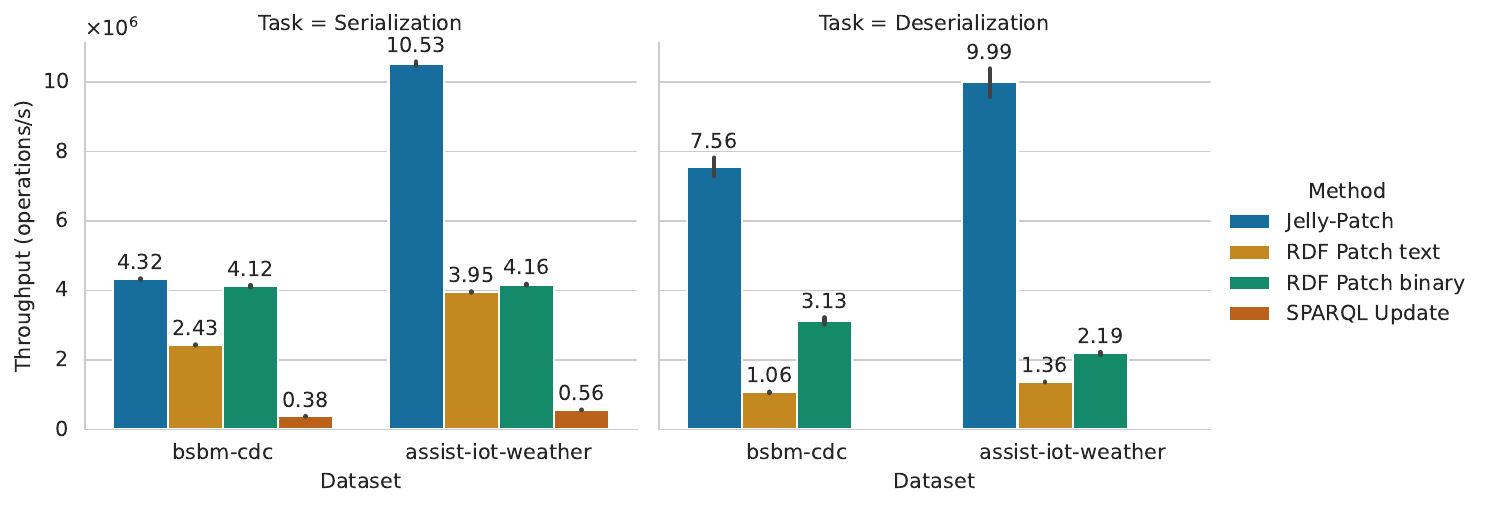}}
    \caption{Serialization and deserialization throughput, in operations per second. Colored bars indicate the mean value. Gray error bars indicate the 95\% confidence interval.}
    \label{fig:ser_des}
\end{figure}

Figure~\ref{fig:ser_des} presents the serialization and deserialization throughput results. SPARQL Update does not have a RDF Patch-compatible parser, so only its serialization speed was tested, with it being by far the slowest format. In \texttt{bsbm-cdc}, Jelly-Patch is only $\sim$5\% faster at serializing than Jena's binary format. This is due to the dataset being a pessimistic case for Jelly-Patch, with a lot of incompressible data (long strings) and few repeating structures. Nonetheless, Jelly-Patch delivers over \textbf{4x} better compression at nearly the same serialization throughput, making it much more advantageous. In \texttt{assist-iot-weather}, Jelly-Patch has \textbf{2.5x} faster serialization than Jena's binary format.

The performance difference is much more pronounced for deserialization. While Jena's binary format was specifically designed to be fast to parse, Jelly-Patch achieves 2.4x higher parsing throughput in \texttt{bsbm-cdc} and \textbf{4.6x} in \texttt{assist-iot-weather}.

It should be noted that Jelly-Patch is purely a new serialization format for RDF Patch, and therefore any benefits it may provide are in the serialization/deserialization speeds and the compression of patches. The internal mechanisms of RDF systems for calculating diffs and applying them on RDF datasets remain the same as with other RDF Patch formats. The significance of the performance impact of using Jelly-Patch in practice will largely depend on the specific systems and their bottlenecks.

\section{Conclusion and Future Work}

In this work, we introduce Jelly-Patch, a fast binary format for recording changes in RDF datasets. In the conducted experiments, it achieved 3.5--8.9x better compression than other RDF Patch formats, while being up to 2.5x faster to serialize, and up to 4.6x faster to parse. This is a significant difference that is expected to greatly benefit large-scale RDF systems, especially in networked or distributed settings. The format is not only useful in CDC scenarios, but also for representing streams of sensor data, achieving 5.4x better compression than the already well-compressed Jelly-RDF.

We plan to continue optimizing both Jelly-RDF and Jelly-Patch through new compression schemes and improvements in implementation efficiency. Also planned are: implementations for Python and Rust, conformance test cases, and extending the \texttt{jelly-cli} with commands for Jelly-Patch. Extensive benchmarks should be performed, using more datasets, which can be contributed to RiverBench~\cite{sowinski2024realizing}.

This work was financially supported from the European Funds under the Sector Agnostic path of the Huge Thing Startup Booster program (project no. 0021/2025, program FENG.02.28-IP.02-0006/23).

\textbf{Declaration on Generative AI.} During the preparation of this work, the authors used ChatGPT in order to draft content. After using this tool, the authors reviewed and edited the content as needed and take full responsibility for the publication’s content.

\vspace{0.3cm}
\noindent
\textbf{Jelly-Patch specification:} \url{https://w3id.org/jelly/dev/specification/patch} \\
\textbf{Jelly-JVM documentation:} \url{https://w3id.org/jelly/jelly-jvm} \\
\textbf{Datasets and full results:} \url{https://doi.org/10.5281/zenodo.16498682}~\cite{zenodo} \\
\textbf{Community contribution guide:} \url{https://w3id.org/jelly/dev/contributing}

%%
%% Define the bibliography file to be used
\bibliography{bibliography}

\end{document}